\documentstyle[]{article}
\font\cero=cmss10 scaled 1728 \font\uno=cmssbx10 scaled 1200
\begin{document}
\begin{flushleft}
{\cero Topological terms and the global symplectic geometry of the
phase space in string theory} \\[3em]
\end{flushleft}
{\sf R. Cartas-Fuentevilla and A. Escalante}\\
{\it Instituto de F\'{\i}sica, Universidad Aut\'onoma de Puebla,
Apartado postal J-48 72570, Puebla Pue., M\'exico.}  \\[4em]

Using an imbedding supported background tensor approach for the
differential geometry of an imbedded surface in an arbitrary
background, we show that the topological terms associated with the
inner and outer curvature scalars of the string worldsheet, have a
dramatic effect on the global symplectic geometry of the phase
space of the theory. By identifying the global symplectic
potential of each Lagrangian term in the string action as the
argument of the corresponding pure divergence term in a
variational principle, we show that those topological terms
contribute explicitly to the symplectic potential of any action
describing strings, without modifying the string dynamics and the
phase space itself. The variation (the exterior derivative on the
phase space) of the symplectic potential generates the integral
kernel of a covariant and gauge invariant symplectic structure for
the theory, changing thus the global symplectic geometry of the
phase space. Similar results for non-Abelian gauge theories and
General Relativity are briefly
discussed.\\

\noindent {\uno I. INTRODUCTION} \vspace{1em}

As everyone knows, within the context of the variational principle
in physics, any term that can be written as a pure surface
divergence does not have effects on the dynamics of the system
under study, and therefore it is common to ignore completely such
terms, and to focus our attention on the equations of motion and
their consequences. For example, in the case of geometrical
theories, it is well known that the Hilbert-Einstein action for
the metric leads to a term proportional to the called Einstein
tensor at the level of the field equations, plus the corresponding
pure divergence term; such field equations correspond to the
Einstein equations in a four-dimensional geometry, but does not
give dynamics to the metric in a two-dimension worldsheet swept
out by a string, since the symmetries of the curvature tensor
imply that the Einstein tensor vanishes for such a geometry.
Therefore, the Einstein-Hilbert action depends, in string theory,
only on the topology of the worldsheet, contributing just with a
pure divergence term, which is of course completely ignored in a
conventional dynamics analysis.

Thus, there is no {\it apparently} any physical motivation for
including such a topological term in any action describing
strings, since the corresponding dynamics remains unaltered. For
example, if we attempt to construct a (conventional) canonical
formulation to quantize the Dirac-Nambu-Goto (DNG) strings from
the corresponding classical dynamics, we shall obtain the same
results whether we include the topological term, which turns out
to be weird, at least from our particular point of view. On the
other hand, it is very known also that the topological term has a
global significance in the path integral formulation of string
theory, weighting the different topologies in the sum over world
surfaces. Thus, it is reasonable to think that a term that depends
only on the global properties of the worldsheet, plays a non
trivial role in such a global description of the theory.

With these preliminaries, the purpose of this work is to show that
in a global description of the canonical formulation of string
theory (as apposed to the {\it local} conventional description of
the canonical formalism in terms of {\it p's} and {\it q's} widely
disseminated in the literature), a topological term has
effectively a global significance, such as it does in the path
integral formulation of the theory. Such a global contribution of
the topological term comes from the argument of the corresponding
pure divergence term in a variational principle, which will be
identified as a global 1-form on the covariant phase space of the
theory, whose direct exterior derivation generates the integral
kernel of a covariant and gauge invariant symplectic structure. As
a by-product, it is shown that from a spurious total divergence
term in a variational principle, one can identify physically
relevant geometrical structures on the phase space.

In the next section, we summarize the basic aspects of the
strongly covariant description of an imbedding given by Carter in
\cite{1}, which are essential for our present aims. In Section
III, we outline the definition of the covariant phase space and
the exterior calculus associated with it. In Section IV, we give
some remarks on the covariant canonical formulation of the DNG
branes, in order to prepare the background for the subsequent
sections, where the topological terms are worked out. Specifically
in Section V, the inner curvature scalar for an imbedded surface
of arbitrary dimension is considered, and we show that in the
particular case of a string world surface, it has not effectively
any contribution to the string dynamics, but we can identify a
symplectic potential for it. In Section VI, we considered again
the inner curvature scalar, but now directly as a pure divergence
term for a two-dimensional geometry; we find a full agreement with
the previous results of Section V. In Section VII, the outer
curvature scalar is considered as a Lagrangian term for string
theory, and we identify for such a topological term its
corresponding symplectic potential. In Section VIII, we give some
concluding remarks, and we discuss some open questions for further
research. In the Appendix A, we discuss the cases of the
Yang-Mills theory and General Relativity and their respective
topological terms. Finally in the Appendix B we summarize the
basic formal aspects of symplectic geometry.
\\[2em]

\noindent {\uno II. Basic differential geometry of an imbedded}
\vspace{1em}

In this Section we introduce the basic ideas of the imbedding
supported background tensor approach developed by Carter \cite{1},
which will be used in the present treatment. In the Carter scheme,
the emphasis is on the use of local coordinate patches on the
background manifold for describing an imbedded p (brane world)
surface in such a higher-dimensional background. The great virtue
of this scheme is that avoids the (widely disseminated) use of
excess mathematical baggage that obscures the simplicity and
generality of laws and results on the subject, which is also
manifested in the study of the symplectic geometry of the brane
dynamics  in \cite{2,3,4}.

Therefore, we outline the description given in \cite{1} for the
various kinds of curvature that are associated with a spacelike or
timelike p-surface imbedded in a n-dimensional space or spacetime
background with metric $g_{\mu\nu}$. Specifically the {\it
internal curvature tensor} of the imbedding can be written as
\begin{equation}
     R_{\kappa\lambda} {^{\mu}}_{\nu} = 2 n_{\sigma} {^{\mu}} \ n_{\nu}
     {^{\tau}} \ n_{[\lambda} {^{\pi}} \ \overline{\nabla}_{\kappa ]}
     \ \rho_{\pi} {^{\sigma}}_{\tau} + 2 \rho_{[\kappa} {^{\mu\pi}}
     \ \rho_{\lambda ]\pi\nu},
\end{equation}
where $n^{\mu\nu}$ is the (first) {\it fundamental tensor} of the
p-surface, that together with the complementary orthogonal
projection $\bot^{\mu\nu}$ satisfy
\begin{equation}
     n^{\mu} {_{\nu}} + \bot^{\mu} {_{\nu}} = g^{\mu} {_{\nu}},
     \quad n^{\mu} {_{\nu}} \ \bot^{\nu} {_{\rho}} = 0,
\end{equation}
and the tangential covariant differentiation operator is defined
in terms of the fundamental tensor as
\begin{equation}
     \overline{\nabla}_{\mu} = n^{\rho} {_{\mu}} \ \nabla_{\rho},
\end{equation}
where $\nabla_{\rho}$ is the usual Riemannian covariant
differentiation operator associated with $g_{\mu\nu}$.
Additionally, $\rho_{\lambda} {^{\mu}}_{\nu}$ represents the
internal frame rotation (pseudo-) tensor field, or more
specifically the background spacetime components of the internal
frame components of the natural gauge connection for the group of
p-dimensional internal frame rotations. The frame gauge dependence
of this field will be crucial in order to establish our pretended
results. It satisfies the properties
\begin{equation}
     \rho_{\lambda\mu\nu} = - \rho_{\lambda\nu\mu}, \quad
     \bot^{\rho} {_{\lambda}} \ \rho_{\rho\mu\nu} = 0 =
     \bot^{\rho} {_{\lambda}} \ \rho_{\mu\rho\nu},
\end{equation}
whereas the internal curvature tensor (1) satisfies the usual
Riemann symmetry properties and the Ricci contractions
\begin{equation}
     R_{\mu\nu} = R_{\mu\sigma\nu} {^{\sigma}}, \quad R =
     R_{\sigma} {^{\sigma}},
\end{equation}
with
\begin{equation}
     \bot^{\sigma} {_{\beta}} \ R_{\sigma\lambda\mu\nu} = 0,
     \quad  \bot^{\sigma} {_{\beta}} \ R_{\sigma\mu} = 0.
\end{equation}
From the fundamental tensor and the Ricci contractions (5) one can
define the internal adjusted Ricci tensor as
\begin{equation}
     \widetilde{R}_{\mu\nu} \equiv R_{\mu\nu} - \frac{1}{2(p -
     1)} \ R \ n_{\mu\nu},
\end{equation}
where $p$ is the dimension of the imbedded p-surface. As pointed
out in \cite{1}, for the special case $p = 2$ of a two-dimensional
imbedded surface (that applies to string theory, for which this
work is concerned), the adjusted Ricci tensor (7) vanishes
identically:
\begin{equation}
     \widetilde{R}_{\mu\nu} \equiv R_{\mu\nu} - \frac{1}{2} R \
     n_{\mu\nu} = 0.
\end{equation}
The identity (8) will imply, as we shall see below, that the inner
curvature scalar given in (5) can not give any effective
contribution in a variational principle, as already it was
mentioned in the introduction.

Similarly, we have the outer curvature tensor of the imbedded in
terms of the external gauge connection
$\omega_{\lambda}{^{\mu}}_{\nu}$ \cite{1},
\[
     \Omega_{\kappa\lambda}{^{\mu}}_{\nu} = 2\bot_{\sigma}
     {^{\mu}} \ \bot_{\nu} {^{\tau}} \ n_{[\lambda} {^{\pi}}
     \overline{\nabla}_{\kappa ]} \ \omega_{\pi}
     {^{\sigma}}_{\tau} + 2 \omega_{[\kappa} {^{\mu\pi}} \
     \omega_{\lambda ]\pi\nu},
\]
with the restricted symmetries
\begin{equation}
     \Omega_{\mu\nu\rho\sigma} = \Omega_{[\mu\nu] \ [\rho\sigma]},
     \quad \Omega_{\lambda\nu} = \Omega_{\sigma\lambda}
     {^{\sigma}}_{\nu} = 0, \quad \Omega = \Omega^{\lambda}
     {_{\lambda}} = 0,
\end{equation}
and thus the outer curvature tensor is purely Weyl-like, since all
its traces vanish. However, as we shall see in Section VII, one
can even construct a (pseudo-) scalar invariant $\Omega$ for
dimensionally restricted geometries \cite{1}, which will be
related with a topological invariant, the outer analogue of the
well known Gauss-Bonnet invariant associated with the inner
curvature scalar. \\[2em]

\noindent {\uno III. Covariant phase space and the exterior
calculus}
\vspace{1em}

In accordance with \cite{5}, in a given physical theory, {\it the
classical phase space is the space of solutions of the classical
equations of motion}, which corresponds to a manifestly covariant
definition. Based on this definition, the idea of giving a
covariant description of the canonical formalism consists in
describing Poisson brackets of the theory in terms of a symplectic
structure on such a phase space in a covariant way, instead of
choosing {\it p's} and {\it q's}. Strictly speaking, a symplectic
structure is a (non degenerate) closed two-form on the phase
space; hence, for working in this scheme an exterior calculus
associated with the phase space is fundamental. We summarize and
adjust all these basic ideas about the phase space formulation
given in Ref.\ \cite{5} for the case of branes treated here
\cite{2}.

Let $Z$ be the phase space; any (unperturbed) background quantity
such as the background and internal metrics, the projection
tensors, connections, etc., will be associated with zero-forms on
$Z$ (see Appendix B). The Lagrangian deformation $\delta$ acts as
an exterior derivative on $Z$, taking $k$-forms into
$(k+1)$-forms, and it should satisfy the nilpotency property,
\begin{equation}
     \delta^{2} = 0,
\end{equation}
and the Leibniz rule
\[
     \delta (A B) = \delta  AB + (-1)^{A} A \ \delta B.
\]
A differential form $A$ that satisfies $\delta A = 0$, is called
{\it closed}. If the differential form $A$ can be written as the
exterior derivative of another form $B$ (of lower order) $A =
\delta B$, is called {\it exact}. Thus, any exact form is
automatically closed, because of the nilpotency property (10). In
general, the differential forms satisfy the Grassman algebra: $AB
= (-1)^{AB} \ BA$.

In particular, the deformation in the coordinate field $x^{\mu}$
of the background $\xi^{\mu} = \delta \ x^{\mu}$, is the exterior
derivative of the zero-form $x^{\mu}$, and corresponds to an
one-form on $Z$, and thus is an anticommutating object: $\xi^{\mu}
\ \xi^{\lambda} = - \xi^{\lambda} \ \xi^{\mu}$. In according to
(10), $\xi^{\mu}$ will be closed, $\delta \ \xi^{\mu} = \delta^{2}
\ x^{\mu} = 0$, which is evident from the explicit form of $\delta
\ \xi^{\mu}$ given in \cite{6}:
\begin{equation}
     \delta \ \xi^{\mu} = - \Gamma^{\mu}_{\lambda\nu} \
     \xi^{\lambda} \ \xi^{\nu} = 0,
\end{equation}
which vanishes because of the symmetry of the background
connection $\Gamma^{\mu}_{\lambda\nu}$ in its indices $\lambda$
and $\nu$ and the anticommutativity of the $\xi^{\lambda}$ on $Z$.
In Appendix B, we summarize other formal aspects of symplectic
geometry, particularly that associated with an ordinary scalar
field, with the idea of clarifying the basic scheme in the
simplest case. A more complete treatment of the symplectic
geometry for systems with support confined to a lower dimensional
submanifold is given in \cite{7}. However, for our purposes, the
results of this section are sufficient. \\[2em]

\noindent {\uno IV. Global structure of the phase space of DNG
branes from a global symplectic potential}
\vspace{1em}

In this section, before considering the topological terms, we
shall give some remarks on the covariant canonical formulation of
the DNG action for an arbitrary brane developed in \cite{2,3}, in
order to prepare the background and to clarify the panorama for
the subsequent inclusion of the topological terms. It is
convenient to do the general treatment for branes of arbitrary
dimension, and then to consider the particular case of string
theory, which will show the particularities of the later with
respect to the former.

The action for DNG branes in a curved embedding background is
given by \cite{6}
\begin{equation}
     S_{0} = \sigma_{0} \int \sqrt{-\gamma} \ d\overline{S},
\end{equation}
where $\sigma_{0}$ is a fixed parameter, $d\overline{S}$ is the
surface element induced on the world surface by the background
metric. The first order (Lagrangian) variation of $S_{0}$ implies
that \cite{6}
\begin{equation}
     \sigma_{o} \int \sqrt{-\gamma} \ \overline{\nabla}_{\mu}
     (n^{\mu} {_{\nu}} \ \xi^{\nu}) d \overline{S} - \sigma_{0}
     \int \sqrt{-\gamma} \ \xi^{\nu} \ \overline{\nabla}_{\mu}
     (n^{\mu} {_{\nu}}) d\overline{S} = 0,
\end{equation}
From Eq.\ (13) it follows that, modulo a total divergence, the
equations of motion are
\begin{equation}
     \overline{\nabla}_{\mu} \ n^{\mu\nu} = K^{\nu} = 0,
\end{equation}
where $K^{\nu}$ is the {\it trace} of the {\it second fundamental
tensor} defined as $K_{\mu\nu} {^{\rho}} = n^{\lambda} {_{\nu}} \
\overline{\nabla}_{\mu} \ n^{\rho} {_{\lambda}}$, thus $K^{\nu} =
K^{\mu} {_{\mu}}^{\nu}$.

From the equations (14) we can define the fundamental concept in
the global description of the canonical formulation of the theory:
the covariant phase space of DNG branes is the space of solutions
of Eqs.\ (14), and we shall call it $Z$.

Following the spirit of the present work of that the pure
divergence term in (13) be no ignored, in \cite{3} we have
demonstrated that the argument of such a term plays the role of a
global symplectic potential on $Z$, in the sense that its exterior
derivative on $Z$ (identified with the deformation operator
$\delta$, according to section III), generates the integral kernel
(the symplectic current) of a (non degenerate) closed two-form on
$Z$, which represents {\it the symplectic structure} that contains
all the physical information on the Hamiltonian structure of the
phase space, representing thus a starting point for the study of
the symmetry and quantization aspects of the theory. Specifically
the symplectic structure is given by
\begin{equation}
     \omega = \int_{\Sigma} \delta (- \sqrt{-\gamma} \ n^{\mu}
     {_{\alpha}} \ \xi^{\alpha}) d \overline{\Sigma}_{\mu} =
     \int_{\Sigma} \sqrt{-\gamma} \ \widetilde{J}^{\mu} \ d
     \overline{\Sigma}_{\mu},
\end{equation}
with $\sqrt{-\gamma} \ \widetilde{J}^{\mu} = \delta
(-\sqrt{-\gamma} \ n^{\mu} {_{\alpha}} \ \xi^{\alpha})$, $\Sigma$
being a (spacelike) Cauchy surface for the configuration of the
brane, and $d \overline{\Sigma}_{\mu}$ is the surface measure
element of $\Sigma$, and is normal on $\Sigma$ and tangent to the
world-surface; Eq.\ (15) shows that $\omega$ is an {\it exact}
differential form (since comes from the exterior derivative of an
one-form), and in particular an identically closed two-form on
$Z$. The closeness is equivalent to the Jacobi identity that
Poisson brackets satisfy, in an usual Hamiltonian scheme.
Moreover, in \cite{3} it is proved that the symplectic current is
(world surface) covariantly conserved ($\overline{\nabla}_{\mu} \
\widetilde{J}^{\mu} = 0$), which guarantees that $\omega$ in (15)
is independent on the choice of $\Sigma$ and, in particular, is
Poincar\'e invariant.

In this manner, as a conclusion of this section, the argument of
the pure divergence term in (13) represents a fundamental (gauge)
field on $Z$, generating the (strength) geometrical structure
$\omega$ on $Z$, which in turns represents a complete {\it
Hamiltonian} description of the covariant phase space of the
theory. \\[2em]

\noindent {\uno V. Inner curvature scalar as a Lagrangian term}
\vspace{1em}

In this section, we shall consider a Hilbert term, proportional to
the inner curvature scalar of the imbedded p-surface
\begin{equation}
     \chi = \sigma_{1} \int \sqrt{-\gamma} \ R \
     d\overline{\Sigma},
\end{equation}
where $\sigma_{1}$ is a fixed parameter, and let us determine its
contribution to the brane dynamics {\it modulo a pure divergence
term} in a variational principle.

Within the covariant scheme given by Carter \cite{6} for the
deformations dynamics, it is known that
\begin{equation}
     \delta \sqrt{-\gamma} = \frac{1}{2} \sqrt{-\gamma} \
     n^{\mu\nu} \ \delta \ g_{\mu\nu},
\end{equation}
where the variation of the background metric is given by its Lie
derivative with respect to the deformation vector field $\xi^{\mu}
= \delta \ x^{\mu}$:
\[
     \delta \ g_{\mu\nu} = \nabla_{\mu} \ \xi_{\nu} + \nabla_{\nu} \
     \xi_{\mu}.
\]
In order to determine the variation of the scalar $R$, let us
calculate first the variation of the internal curvature tensor
(1), and its contractions, exploiting the frame gauge dependence
of $\rho_{\lambda}{^{\mu}}_{\nu}$, which means that it can always
be set equal to zero at any single chosen point by an appropriate
choice of the relevant frames \cite{1}. Therefore, if we consider
a variation of $\rho_{\lambda} {^{\mu}}_{\nu}$ to a new connection
\[
     \rho_{\lambda}{^{\mu}}_{\nu} \rightarrow \rho_{\lambda}
     {^{\mu}}_{\nu} + \delta \ \rho_{\lambda}{^{\mu}}_{\nu},
     \nonumber \\
\]
then this variation leads to a variation in the internal curvature
tensor given by
\begin{equation}
     \delta \ R_{\kappa\lambda} {^{\mu}}_{\nu} = 2 n_{\sigma}
     {^{\mu}} \ n_{\nu} {^{\tau}} \ n_{[\lambda} {^{\pi}} \
     \overline{\nabla}_{\kappa ]} \ \delta \ \rho_{\pi}
     {^{\sigma}}_{\tau},
\end{equation}
and thus,
\begin{equation}
     \delta \ R_{\mu\nu} = 2 n_{\sigma} {^{\kappa}} \ n_{\nu}
     {^{\tau}} \ n_{[\mu} {^{\pi}} \ \overline{\nabla}_{\kappa ]}
     \ \delta \ \rho_{\pi} {^{\sigma}}_{\tau},
\end{equation}
and hence, we find that
\begin{equation}
     n^{\mu\nu} \ \delta \ R_{\mu\nu} = \overline{\nabla}_{\mu} \
     \psi^{\mu}_{top},
\end{equation}
where
\begin{equation}
     \psi^{\mu}_{top} = n^{\alpha\beta} \ \delta \
     \rho_{\alpha}{^{\mu}}_{\beta} - n^{\alpha}_{\beta} \ n^{\mu\tau} \
     \delta \ \rho_{\alpha} {^{\beta}}_{\tau}.
\end{equation}
It is important to note that, although the connection
$\rho_{\alpha} {^{\beta}}_{\tau}$ can always be set equal to zero,
its variation $\delta\rho_{\alpha} {^{\beta}}_{\tau}$, a tensor
field, can no be in general gauged away.

The equations (18)--(20) may be the analogue of the very known
Palatini equations in the context of general relativity, where
such equations are used in order to obtain the Einstein equations
from the Hilbert action.

Finally, using Eqs.\ (18) and (21), and considering that $R =
n^{\mu\nu} \ R_{\mu\nu}$, one can calculate the variation of
$\chi$ in (16):
\begin{equation}
     \delta \ \chi = \sigma_{1} \int \sqrt{-\gamma} \ \left(
     \frac{1}{2} R \ n^{\mu\nu} - R^{\mu\nu} \right) \delta
     g_{\mu\nu}d\overline{\Sigma} + \sigma_{1} \int \sqrt{-\gamma}\overline{\nabla}_{\mu} \
     \psi^{\mu}_{top}d\overline{\Sigma};
\end{equation}
Eq.\ (22) gives the universal contribution of the (inner)
curvature scalar as a Lagrangian term on the brane dynamics
through the first term on the right hand side, and, as one can
already guess at this point, its universal contribution to the
symplectic potential on the phase space of the brane theory
through $ \psi^{\mu}_{top}$, in the second term. In general,
$\frac{1}{2} R \ n^{\mu\nu} - R^{\mu\nu}$ does not vanish for a
geometry of arbitrary dimension, and therefore, in general, $\chi$
change simultaneously the brane dynamics, the phase space, and the
symplectic structure of the later (and hence this work may no make
sense in such a general situation). However, as discussed in
Section II (see Eq.\ (8)), the adjusted Ricci tensor vanishes
identically for string theory, and $\chi$ does not give dynamics
to such objects (and for convenient boundary conditions, the
surface term in (22) can be completely eliminated). Moreover, by
defining the covariant phase space as the space of solutions of
the dynamics equations, the phase space itself is unmodified by
the inclusion of $\chi$ in string theory. However, considering our
present procedure for identifying the contribution of any
Lagrangian term to the global symplectic potential of the theory,
$\chi$ in Eq.\ (22) has already modified the symplectic structure
of the (unmodified) phase space of the string theory by means of $
\psi^{\mu}_{top}$. For instance, if we consider the more general
action $S = S_{0} + \chi$ for the particular case of string
theory, where $S_{0}$ and $\chi$ are given in Eqs.\ (12) and (16),
respectively, the equations of motion are, according to Eqs.\ (8),
(13), and (22), the same equations (14) obtained just for $S_{0}$,
and hence the phase space is again $Z$. However, the symplectic
potential on $Z$ is no longer $\sigma_{0} \ n^{\mu}_{\nu} \
\xi^{\nu}$, but $\sigma_{0} \ n^{\mu}_{\nu} \ \xi^{\nu} +
\sigma_{1} \ \psi^{\mu}_{top}$, and the corresponding symplectic
structure is given by
\begin{equation}
     \omega = \int_{\Sigma} \delta \left[ \sqrt{-\gamma} \
     (\sigma_{0} \ n^{\mu}_{\nu} \ \xi^{\nu} + \sigma_{1} \
     \psi^{\mu}_{top}) \right] d \ \overline{\Sigma}_{\mu},
\end{equation}
which will be evidently closed, and similarly for any action
describing strings. \\ [2em]

\noindent {\uno VI. The inner curvature scalar for a string
worldsheet}
\vspace{1em}

In the previous section we have found the contribution of the
Gauss-Bonnet topological term to the Hamiltonian structure of
string theory considering the more general case of a brane of
arbitrary dimension, which shown the particularities of the string
case as opposed to the other higher-dimensional objects. However,
we can determine the symplectic potential for string theory
directly from the expression for the inner curvature scalar for a
two-dimensional worldsheet, which has the well known property of
being a pure surface divergence, avoiding the general brane
geometry, and exploiting the particularities of a two-dimensional
geometry.

In the strong covariant scheme given in \cite{1}, it is shown that
the inner curvature scalar can be written as
\begin{equation}
     R = \overline{\nabla}_{\mu} \ ({\cal E}^{\mu\nu} \ \rho_{\nu}),
\end{equation}
where the frame independent antisymmetric unit surface element
tensor ${\cal E}^{\mu\nu}$ is defined as
\begin{equation}
     {\cal E}^{\mu\nu} = 2 \ \iota_{0}{^{[\mu}} \ \iota_{1}{^{\nu ]}};
\end{equation}
$\iota_{0}^{\mu}$ is a timelike unit vector, and $\iota_{1}^{\mu}$
a spacelike  one, which constitute an orthonormal tangent (to the
worldsheet) frame. The rotation (co)vector $\rho_{\mu}$ is defined
in terms of the internal connection as
\begin{equation}
     \rho_{\lambda} = \rho_{\lambda}{^{\mu}}_{\nu} \ {\cal
     E}^{\nu}{_{\mu}}, \quad \rho_{\lambda} {^{\mu}}_{\nu} =
     \frac{1}{2} \ {\cal E}^{\mu}{_{\nu}} \ \rho_{\lambda}.
\end{equation}
In accordance with Eqs.\ (26), the frame gauge dependence of
$\rho_{\lambda}{^{\mu}}_{\nu}$ induces the same gauge dependence
on $\rho_{\mu}$ (see paragraph after Eq.\ (17)); therefore a
variation $\rho_{\mu} \rightarrow \rho_{\mu} + \delta \
\rho_{\mu}$ leads to a variation in (frame gauge dependent) $R$
given by
\begin{equation}
     \delta \ R = \overline{\nabla}_{\mu} \ ({\cal E}^{\mu\nu} \ \delta \
     \rho_{\nu}).
\end{equation}
Hence, the variation of $\chi$ in Eq.\ (16) (with $R$ given by
(24) for a string) can be written simply as
\begin{equation}
     \delta \ \chi = \sigma_{1} \int \sqrt{-\gamma} \
     \overline{\nabla}_{\mu} \ ({\cal E}^{\mu\nu} \ \delta \
     \rho_{\nu}) \ d \overline{\Sigma},
\end{equation}
and we identify  $(\sigma_{1}) {\cal E}^{\mu\nu} \delta
\rho_{\nu}$ as a symplectic potential for $\chi$ in string theory.
Considering that from Eqs.\ (26), $\delta \
\rho_{\lambda}{^{\mu}}_{\nu} = \frac{1}{2} {\cal E}^{\mu}{_{\nu} \
\delta \rho_{\lambda}}$, that $n^{\mu\nu} = \iota_{0}^{\mu} \
\iota^{0\nu} + \iota_{1}^{\mu} \ \iota^{1\nu}$ \cite{1}, and Eq.\
(25), it is very easy to verify that $\psi^{\mu}_{top}$ in Eq.\
(21) corresponds, for string theory, exactly to ${\cal E}^{\mu\nu}
\delta \rho_{\nu}$. In this manner, there exists a full agreement
between both approaches for finding out the symplectic potential
for $\chi$. Note that in this case, there is no
restriction on the dimension of the background geometry. \\[2em]

\noindent {\uno VII. The outer curvature scalar for a string
worldsheet}
\vspace{1em}

In the case of a world sheet embedded in a four-dimensional
background spacetime, using the standard fully antisymmetric
four-volume measure tensor of the background
${\varepsilon}^{\lambda\mu\nu\rho}$, and the outer curvature
scalar given in Section II, one can determine a scalar magnitude
$\Omega$ given by [1, see also 6, and 8]
\[
     \Omega = \frac{1}{2} \ \Omega_{\lambda\mu\nu\rho} \
     {\varepsilon}^{\lambda\mu\nu\rho},
\]
and a {\it twist} convector $\omega_{\mu}$ (the outer analogue of
$\rho_{\mu}$), in the form
\begin{equation}
     \omega_{\nu} = \frac{1}{2} \ \omega_{\nu} {^{\mu\lambda}} \
     {\varepsilon}_{\lambda\mu\rho\sigma} \ {\cal E}^{\rho\sigma},
\end{equation}
and therefore, we can rewrite $\Omega$ as a pure divergence as
\begin{equation}
     \Omega = \overline{\nabla}_{\mu} \ ({\cal E}^{\mu\nu} \
     \omega_{\nu}),
\end{equation}
which is frame gauge dependent and is the (dimensionally
restricted) outer analogue of $R$ in Eq.\ (24). Thus, the world
surface integral of $\Omega$ gives a topological term expressed as
\begin{equation}
     \chi' \equiv \sigma_{2} \int \sqrt{-\gamma} \ \Omega \
     d\overline{\Sigma} = \sigma_{2} \int \sqrt{-\gamma} \
     \overline{\nabla}_{\mu} \ ({\cal E}^{\mu\nu} \
     \omega_{\nu}) \ d \overline{\Sigma},
\end{equation}
where $\sigma_{2}$ is a fixed parameter. Hence, $\chi'$ gives no
an effective contribution on the string dynamics. Appealing to the
frame gauge dependence of $\omega_{\nu}$ inherited from
$\omega_{\nu}{^{\mu\lambda}}$, the variation of $\chi'$ is given
by
\begin{equation}
     \delta \ \chi' = \sigma_{2} \int \sqrt{-\gamma} \
     \overline{\nabla}_{\mu} \ ({\cal E}^{\mu\nu} \ \delta \
     \omega_{\nu}) \ d \overline{\Sigma},
\end{equation}
and we identify $(\sigma_{2}) {\cal E}^{\mu\nu} \ \delta \
\omega_{\nu}$ as a symplectic potential for $\chi'$ in string
theory (in a four-dimensional background). Note that although
$\omega_{\nu}$ can be set equal to zero at any single point,
$\delta \ \omega_{\nu}$ can be no in general gauged away. \\[2em]

\noindent {\uno VIII. Remarks and prospects}
\vspace{1em}

\noindent \textbf{A. Topological terms and deformation dynamics}

An important conclusion from the previous sections is that the
topological terms do not need modify the equations of motion of
the theory for having an effective contribution on the symplectic
properties of the phase space. Furthermore, the procedure followed
in the present work for determining the corresponding symplectic
potential for any Lagrangian term, may seem only a prescription
without any solid basis (although there no either exists
apparently some argument for ignoring such contributions), with
the final purpose that the variation of those potentials will give
an effective contribution on the integral kernel of the symplectic
structure.

However, we have employed another approach for determining the
contribution of any Lagrangian term on the kernel integral of the
symplectic structure of the theory under study, and it consists in
to construct the symplectic current from the corresponding
deformation dynamics, using the concept of adjoint operators. For
example, in \cite{9} the symplectic currents originally suggested
in \cite{5} for Yang-Mills theory and General Relativity were
found using the adjoint operators scheme. The corresponding
current for branes in a curved background was obtained also using
such a scheme in \cite{2}; such a current corresponds exactly to
that obtained using the procedure presented here, as we have
outlined in Section IV. All these results suggest clearly the
following: the topological terms can change drastically the
deformation dynamics of the theory, without modifying the dynamics
itself. With this idea, in \cite{10} it is proved that the
variation of the Einstein tensor modifies the deformation dynamics
of string theory in a weakly (as opposed to the present strongly)
covariant description of the theory; the symplectic current
obtained from that modified deformation dynamics using the adjoint
operators scheme, corresponds exactly to that obtained by
variations of the symplectic potential obtained in \cite{11} using
the present procedure. Therefore, it remains a detailed study of
the contributions of the topological terms considered in the
present work, on the deformation dynamics of string theory. The
importance of the study of a (modified) deformation dynamics goes
beyond the present interest in the symplectic geometry of the
phase space, for example the stability aspects of the solutions of
the theory, among other. \\

\noindent \textbf{B. Possible physical implications of
$\psi^{\mu}_{top}$}

Once we have determined explicitly the contribution of the
topological terms on the global phase space formulation for string
theory, it is important to discuss about the possible implications
in a more physical context. As discussed in the introduction, a
symplectic structure on the phase space is finally a {\it
Hamiltonian} structure for the theory, and thus represents a
starting point for the study of the symmetry and quantization
aspects. For example, in \cite{4} the Poincar\'{e} charges, the
closeness of the Poincar\'{e} algebra, relevant commutation
relations for DNG branes (which contains the string case as a
particular case) were studied using precisely the symplectic
geometry of the phase space established in \cite{2,3}. Does change
the inclusion of the topological terms the results obtained in
\cite{4} for DNG strings, leaving the dynamics unchanged?

\noindent \textbf{C. A new type of topological strings?}

The Lagrangian terms associated with the topological terms always
have been considered as corrective or additional terms of other
Lagrangian terms, which whether give dynamics to the system.
Apparently the fact that the topological terms leave unchanged the
dynamics does no permit that one may consider only such terms
``for making physics". However, the results presented here may
suggest that, in spite of its null dynamics, the existence of
$\psi^{\mu}_{top}$ for a Lagrangian involving only the topological
terms may imply that ``the physics" for such (hypothetical) {\it
topological strings} will be in other domain, different to the
classical one, since finally a symplectic structure (that obtained
from $\psi^{\mu}_{top}$) governs the ``transition" between the
classical and quantum domains. However, it is important to point
out that we are only speculating. It is possible that finally the
answer for this open question turns out
to be trivially simply. \\

\noindent \textbf{D. Final comments}

The results presented here may extend the role played by the
topological terms in the context of string theory, and possibly
other areas. Such results, although certainly limited, suggest a
deeper research with a perspective different to that usually given
in the known literature.

As we have seen in this work, the differential geometry of an
imbedded developed in \cite{1} has been crucial in order to
establish our results, and various aspects considered in that
reference had not been completely treated even in the pure
mathematical context, as pointed out by Carter himself \cite{1}.
In this sense, following the spirit of the mathematical physics,
the present work can be considered as an attempt for extracting
{\it physics} from the Carter formalism. \\[2em]

\begin{center}
{\uno ACKNOWLEDGMENTS}
\end{center}
\vspace{1em}

The work of A. E. was supported by CONACYT (M\'exico), and R.C.F.
acknowledges the support by the Sistema Nacional de Investigadores
(M\'exico). The authors wish to thank the editors for the
invitation to participate in this project, particularly F.
Columbus. The authors also thank  F. Rojas for typing the original
manuscript. \\[2em]

\begin{center}
{\uno Appendix A: Symplectic potentials and topological terms for
Yang-Mills theory and General Relativity}
\end{center}
\vspace{1em}
\renewcommand{\theequation}{A\arabic{equation}}
\setcounter{equation}{0} \vspace{1em}

\noindent \textbf{A. Yang-Mills Theory}

As we know, the Yang-Mills theory is a generalization of Maxwell's
electromagnetic theory, and the action for this theory is
\begin{equation}
     L = -\frac{1}{2} \int Tr( F^{\mu\nu} \ F_{\mu \nu})d^{4}x,
\end{equation}
where $F_{\mu\nu}$ is the Yang-Mills curvature given by
\begin{equation}
     F_{\mu \nu} = \partial_{\mu} \ A_{\nu} - \partial_{\nu} \
     A_{\mu} + [A_{\mu},A_{\nu}],
\end{equation}
and $A_{\nu}$ is the gauge connection. The variation of the
curvature (A2) is
\begin{equation}
     \delta \ F_{\mu \nu} = \partial_{\mu} \ \delta \ A_{\nu} -
     \partial_{\nu} \ \delta \ A_{\mu} + [\delta \ A_{\mu},A_{\nu}]
     + [A_{\mu},\delta \ A_{\nu}].
\end{equation}
In this manner, using the equation (A3) we can calculate the
variation of the action (A1), as usual in ordinary field theory,
and to obtain
\begin{eqnarray}
     \delta \ L \!\! & = & \!\! - \int Tr(F^{\mu\nu} \
     \delta \ F_{\mu\nu})d^{4}x \nonumber \\
     \!\! & = & \!\! - 2 \int \partial_{\mu} Tr[(\delta \
     A_{\nu} \ F^{\mu\nu})]d^{4}x + 2 \int Tr[(\partial_{\mu} \
     F^{\mu\nu}+ [A_{\mu},F^{\mu\nu}]) \delta \ A_{\nu}]d^{4}x,
\end{eqnarray}
where we can find the equations of motion
\begin{equation}
     \partial_{\mu} \ F^{\mu\nu} + [A_{\mu},F^{\mu\nu}] = 0,
\end{equation}
and following the ideas of the present work, the argument of the
pure divergence term in (A4) (traditionally ignored in the
literature),
\begin{equation}
     \Psi^{\mu} \equiv - Tr[\delta \ A_{\nu} \ F^{\mu\nu}],
\end{equation}
that does not contribute to the dynamics of the system, works as a
symplectic potential on phase space.

If we take the variation of $\Psi^{\mu}$ in equation (A6), we
obtain
\begin{equation}
     \delta \ \Psi^{\mu} = Tr[\delta \ A_{\nu} \ \delta \
     F^{\mu\nu}],
\end{equation}
where we have considered that $\delta$ is nilpotent and the
Leibniz rule. We can see that the expression (A7) is exactly the
symplectic current suggested in \cite{5} and obtained in \cite{9}
applying the method of self-adjoint operators. Thus, $\delta \
\Psi^{\mu}$ is a covariantly conserved because of the
self-adjointness of the linearized theory \cite{9}, and by
equation (A7) is closed because of the nilpotency of $\delta$.
Therefore, the two-form
\begin{equation}
     \omega = \int_{\Sigma} \delta (\Psi^{\mu}) d \Sigma_{\mu},
\end{equation}
where $\Sigma$ is a Cauchy hypersurface, is a symplectic structure
for Yang-Mills theory.

Furthermore, it is well known that one can construct a topological
term for the Yang-Mills theory given essentially by
\begin{equation}
     \epsilon^{\mu\nu\sigma\rho} \ Tr(F_{\mu\nu} \
     F_{\sigma\rho}),
\end{equation}
which is a total derivative and does not give dynamics to the
gauge field, but as we already known, will contribute explicitly
to the symplectic potential $\Psi^{\mu}$ in Eq.\ (A6) (and to the
linearized equations), without modifying the equations of motion
(A5). \\[2em]

\noindent \textbf{B. General Relativity}

Let us  consider the Einstein-Hilbert action
\begin{equation}
     L = \int \sqrt{-g} \ R \ d^{4}x,
\end{equation}
where $g$ is the determinant of the metric tensor, and $R$ is the
scalar curvature. Considering that
\begin{equation}
     \delta \ \sqrt{-g} = \frac{1}{2} \sqrt{-g} \ g^{\mu\nu} \
     \delta \ g_{\mu\nu},
\end{equation}
\begin{equation}
     \delta \ g^{\mu\nu} = - g^{\mu\alpha} \ g^{\gamma\nu} \
     \delta \ g_{\alpha\gamma},
\end{equation}
\begin{equation}
     \delta \ R_{\mu\nu} = \nabla_{\gamma} \ \delta \
     \Gamma{_{\mu\nu}}^{\gamma} - \nabla_{\nu} \ \delta \
     \Gamma{_{\mu\gamma}}^{\gamma},
\end{equation}
we can calculate the variation of $R$ using the equations (A12)
and (A13), obtaining
\begin{eqnarray}
     \delta R \!\! & = & \!\! \delta (g^{\mu\nu} \ R_{\mu\nu}) =
     \delta \ g^{\mu\nu} \ R_{\mu\nu} + g^{\mu\nu} \ \delta \
     R_{\mu\nu} \nonumber \\
     \!\! & = & \!\! - g^{\mu\alpha} \ g^{\gamma\nu} \ \delta \
     g_{\alpha\gamma} \ R_{\mu\nu} + g^{\mu\nu}[\nabla_{\gamma} \
     \delta \ \Gamma{_{\mu\nu}}^{\gamma} - \nabla_{\nu} \delta \
     \Gamma{_{\mu\gamma}}^{\gamma}].
\end{eqnarray}
In this manner, the variation of $L$ in equation (A10) is
\begin{eqnarray}
     \delta L \!\! & = & \!\! \int [\delta \sqrt{-g} \ R +
     \sqrt{-g} \ \delta R]d^{4}x \nonumber \\
     \!\! & = & \!\! \int \sqrt{-g}(- R^{\mu\nu} + \frac{1}{2}
     g^{\mu\nu} \ R) \delta \ g_{\mu\nu}d^{4}x + \int \sqrt{-g}
     \nabla_{\gamma} [g^{\mu\nu} \ \delta \
     \Gamma{_{\nu\mu}}^{\gamma} - g^{\mu\gamma} \ \delta \
     \Gamma{_{\mu\alpha}}^{\alpha}] d^{4}x,
\end{eqnarray}
where we can identify the very known equations of motion
\begin{equation}
     R^{\mu\nu}- \frac{1}{2} g^{\mu\nu} \ R = 0,
\end{equation}
and we identify from the pure divergence term in equation (A15),
the following
\begin{equation}
     \Psi^{\gamma} = \sqrt{-g}[g^{\mu\nu} \ \delta
     \Gamma{_{\nu\mu}}^{\gamma} - g^{\mu\gamma} \ \delta
     \Gamma{_{\mu\alpha}}^{\alpha}]
\end{equation}
as a symplectic potential for General Relativity, that does not
contribute to the dynamics of system but will generate a
symplectic structure on the phase space.

If we take the variation of (A17) we find
\begin{eqnarray}
     \delta \ \Psi^{\gamma} \!\! & = & \!\! \delta \ \sqrt{-g}
     [g^{\mu\nu} \ \delta \ \Gamma{_{\nu\mu}}^{\gamma} -
     g^{\mu\gamma} \ \delta \ \Gamma{_{\mu\nu}}^{\nu}] + \sqrt{-g}
     \ \delta[g^{\mu\nu} \ \delta \ \Gamma{_{\nu\mu}}^{\gamma} -
     g^{\mu\gamma} \ \delta \ \Gamma{_{\mu\nu}}^{\nu}] \nonumber
     \\
     \!\! & = & \!\! - \sqrt{-g} \ j^{\gamma},
\end{eqnarray}
where $j^{\gamma}$ is given by
\begin{equation}
     j^{\gamma} = \delta \ \Gamma{_{\nu\mu}}^{\gamma} [\delta \
     g^{\mu\nu} + \frac{1}{2} g^{\mu\nu} \ \delta \ lng ] - \delta
     \ \Gamma{_{\mu\nu}}^{\nu} [\delta \ g^{\gamma\mu} +
     \frac{1}{2} g^{\gamma\mu} \ \delta \ lng],
\end{equation}
that is the expression suggested in \cite{5} and obtained in
\cite{9} applying the method of self-adjoint operators. In this
manner $j^{\gamma}$ is covariantly conserved because of the
self-adjointness of the linearized theory \cite{9}, and by
equation (A18) is closed.

Therefore
\begin{equation}
     \omega = \int_{\Sigma} \delta (\Psi^{\gamma}) d
     \Sigma_{\gamma},
\end{equation}
is a symplectic structure for General Relativity.

Similarly, in the case of General Relativity one can construct a
topological term given, for specific numbers A, B, and C, as
\begin{equation}
     \epsilon^{\mu\nu\alpha\beta} \ Tr(R_{\mu\nu} \
     R_{\alpha\beta}) = \sqrt{-g}\ \left[ A \ R^{2}_{\mu\nu\alpha\beta}
     + B \ R^{2}_{\alpha\beta} + C \ R^{2} \right],
\end{equation}
which, for convenient boundary conditions does not change the
equations (A16), but it will contribute explicitly to the
symplectic structure of the theory (A20).

It is well known the relevant role that the topological terms
(A9), and (A21) play in the called gauge and gravitational
anomalies (respectively) in the Feynman path integral formulation
of the theories. Therefore, the results presented in this work may
contribute to the study of the profound relationship between pure
divergence terms, and topological numbers, but now in the setting
of a canonical formulation of the theories, which is a subject
practically unknown in the literature. We hope to
extent all these subjects elsewhere. \\[2em]

\begin{center}
{\uno Appendix B: Basic symplectic geometry}
\end{center}
\vspace{1em}
\renewcommand{\theequation}{B\arabic{equation}}
\setcounter{equation}{0} \vspace{1em}

In this appendix we discuss briefly the basic elements of the
symplectic geometry of a scalar field theory on spacetime $(M)$,
with the purpose of clarifying our basic ideas of Section IV of
the present work. For more details about this little outline, see
references \cite{5}.

If
\begin{equation}
     \Delta \phi - V'(\phi) = 0,
\end{equation}
is the standard equation of motion for the scalar field $\phi$ on
spacetime, the covariant phase space $Z$ corresponds in this case
to the space of solutions of Eq.\ (B1). In this manner, if
$\phi\in Z$ and $x\in M$ is a spacetime point, then we can define
a function $\hat{x}$ on $Z$ by the mapping $\hat{x}: Z\rightarrow
R$, $\hat{x}(\phi)=\phi(x)$. The elements  of the tangent vector
space to $Z$ at $\phi$ ($T_{\phi}Z$) correspond to solutions of
the linearized equation
\begin{equation}
     \Delta \delta \phi - V''(\phi) \ \delta\phi = 0,
\end{equation}
where $\delta\phi$ is an infinitesimal displacement of $\phi$.
Furthermore, we can define an one-form $x^{*}$ (an element of the
dual space  $T_{\phi}^{*}Z$ to  $T_{\phi}Z$), by the mapping
$T_{\phi}Z\rightarrow R$, $x^{*}(\delta\phi)=\delta\phi(x)$. In
this manner $\delta$ associates to the zero-form $\hat{x}:
Z\rightarrow R$, the one-form $\delta \hat{x}\equiv x^{*}:
TZ\rightarrow R$, according to the rule
\begin{equation}
      \delta(\phi(x)) = \delta(\hat{x}(\phi)) \equiv \delta
      \hat{x}(\delta\phi) = x^{*}(\delta\phi) = \delta\phi(x).
\end{equation}
Therefore, misusing this definitions, we can denote the function
$\hat{x}(\phi)$ as $\phi(x)$, and the one-form
$\delta\hat{x}(\delta\phi)$ as $\delta\phi(x)$.

A general $n$-form can be written as
\begin{equation}
     A = \int dx_{1}... dx_{n}\alpha_{x_{1}...
     x_{n}}(\phi)\delta\phi(x_{1})...\delta\phi(x_{n}),
\end{equation}
where $\alpha_{x_{1}... x_{n}}(\phi)$ is an arbitrary function for
each $n$-tuple of the spacetime points $x_{1},... ,x_{n}$. The
action of $\delta$ on this $n$-form is defined as
\begin{equation}
     \delta A = \int dx_{0}dx_{1}... dx_{n}\frac{\delta\alpha_{x_{1}...
     x_{n}}}{\delta\phi(x_{0})}\delta\phi(x_{0})\delta
     \phi(x_{1})...\delta\phi(x_{n}),
\end{equation}
where $\frac{\delta\alpha}{\delta\phi}$ denotes the variational
derivative of $\alpha$ with respect to $\phi(x)$. From this
equation, it is easy to see that
\begin{equation}
     \delta^{2} A = \int dx_{0}'dx_{0}dx_{1}...
     dx_{n}\frac{\delta^{2}\alpha_{x_{1}...
     x_{n}}}{\delta\phi(x_{0}') \delta\phi(x_{0})} \delta\phi(x_{0}')
     \delta\phi(x_{0}) \delta\phi(x_{1})... \delta\phi(x_{n})=0,
\end{equation}
which vanishes identically since the variational derivative in
(B6) is symmetric respect to the interchange of $\phi(x_{0}')$ and
$\phi(x_{0})$, and the fact that $\delta\phi(x_{0}')$ and
$\delta\phi(x_{0})$ are anticommutating objects (correspond, as
seen above, to one-forms). Thus, we can establish that, in
general,
\begin{equation}
     \delta^{2} = 0.
\end{equation}
\\[2em]

\end{document}